\documentclass{ws-ijns}
\usepackage{graphicx}
\usepackage[english]{babel}
\usepackage{program}
\usepackage{url}
\usepackage[T1]{fontenc}
\usepackage[ansinew]{inputenc}
\usepackage[caption=false]{subfig}
\usepackage[super,sort,compress]{cite}
\usepackage{colortbl}

\begin{document}


\markboth{Jin Zhu, Chuan Tan, Junwei Yang, Guang Yang, Pietro Lio'}{MIASSR: An Approach for Medical Image Arbitrary Scale Super-Resolution}

\title{MIASSR: An Approach for Medical Image Arbitrary Scale Super-Resolution}

\author{Jin Zhu\footnote{Send correspondence to Jin Zhu (jin.zhu@cl.cam.ac.uk) and Guang Yang (g.yang@imperial.ac.uk).}}
\address{Department of Computer Science and Technology, \\ University of Cambridge, Cambridge, CB3 0FD, United Kingdom}

\author{Chuan Tan}
\address{Department of Computer Science and Technology, \\ University of Cambridge, Cambridge, CB3 0FD, United Kingdom}

\author{Junwei Yang}
\address{Department of Computer Science and Technology, \\University of Cambridge, Cambridge, CB3 0FD, United Kingdom}

\author{Guang Yang\footnote{Guang Yang and Pietro Lio' are senior and last co-authors contributed equally for this study.}}
\address{Cardiovascular Research Centre, \\Royal Brompton Hospital, London, SW3 6NP, United Kingdom
\\National Heart and Lung Institute, \\Imperial College London, London, SW7 2AZ, United Kingdom}

\author{Pietro Lio'}
\address{Department of Computer Science and Technology, \\ University of Cambridge, Cambridge, CB3 0FD, United Kingdom}

\maketitle

\begin{abstract}
Single image super-resolution (SISR) aims to obtain a high-resolution output from one low-resolution image.  Currently, deep learning-based SISR approaches have been widely discussed in medical image processing, because of their potential to achieve high-quality, high spatial resolution images without the cost of additional scans. However, most existing methods are designed for scale-specific SR tasks and are unable to generalise over magnification scales. In this paper, we propose an approach for medical image arbitrary-scale super-resolution (MIASSR), in which we couple meta-learning with generative adversarial networks (GANs) to super-resolve medical images at any scale of magnification in $(1, 4]$. Compared to state-of-the-art SISR algorithms on single-modal magnetic resonance (MR) brain images (OASIS-brains) and multi-modal MR brain images (BraTS), MIASSR achieves comparable fidelity performance and the best perceptual quality with the smallest model size. We also employ transfer learning to enable MIASSR to tackle SR tasks of new medical modalities, such as cardiac MR images (ACDC) and chest computed tomography images (COVID-CT). The source code of our work is also public. Thus, MIASSR has the potential to become a new foundational pre-/post-processing step in clinical image analysis tasks such as reconstruction, image quality enhancement, and segmentation.
\end{abstract}


\begin{multicols}{2}
\section{Introduction}

Medical images of high quality and resolution are important in the current clinical process. For example, features and biomarkers extracted from functional and structural magnetic resonance (MR) images play a key role in the diagnosis and study of Alzheimer's disease \cite{ADDiagnosis}, stroke \cite{Stroke}, autism \cite{Autism}, and Parkinson's disease \cite{Parkinson}. However, the spatial resolution of medical images is limited by constraints such as acquisition time and equipment. Thus, super-resolution (SR) methods, which aim to increase the spatial resolution of digital images as post-processing, have the potential to improve the quality of medical images without additional scanning costs \cite{MedSRReview}. SR methods consider low-resolution (LR) images as the degradation of high-resolution (HR) images and magnify them by approximating a super-resolving transformation \cite{SRMicroscopy, SRReview18, SRReview} to generate new pixels based on LR images and prior knowledge. Depending on the number of input and output images, SR methods can be briefly classified into single image SR and multi-image SR. In this paper, we focus on single image super-resolution (SISR), which recovers one high resolution (HR) image from its low resolution (LR) version. In recent years, deep learning techniques have achieved great success in image processing problems such as human gait recognition \cite{IJNSLSTM} and image anomaly detection \cite{IJNSDetection}. Particularly, with the rapid development of convolutional neural networks (CNNs) and generative adversarial networks (GANs), deep-learning-based SR methods have achieved remarkable performance \cite{LFSR, MSGAN, CycleGANSR, UNetSR} on various medical image modalities. However, most of these works are designed for specific magnification scales and treat SR with different scales as independent tasks. Thus, several models have to be trained and stored for different magnification tasks. Furthermore, it is challenging to collect large clinical datasets of high- and low-resolution image pairs to train these SR methods for new applications. The high cost of training and implementation lead to poor clinical applicability and limit their application as a result. In this work, we first seek to apply meta-learning to GANs to tackle scale-free super-resolution in medical images. We implement an end-to-end medical image SR network, which takes one LR image as input and generates corresponding SR images of an arbitrary scale of magnification. We train this model with pixel-wise error, perceptual loss and GAN based adversarial loss, to improve both the perceptual quality and fidelity of generated images. Moreover, we overcome the cost of modifying well-trained models to new medical modalities using transfer learning. In particular, we focus on SR tasks with arbitrary scales in $(1, 4]$. This range is chosen as it is most commonly used in practice, but larger scales can be tackled using proper training data. The rest of the paper is structured as follows: first, we review the state-of-the-art (SOTA) SISR methods for natural and medical images; second, we describe the proposed method in detail; next, we introduce the design of our experiments, including data, evaluation metrics, and implementation details; and finally we illustrate the results of MIASSR on four medical image datasets and discuss the influence of each component of MIASSR.

\begin{figure*}[t]
	\centering
	\includegraphics[width=\textwidth]{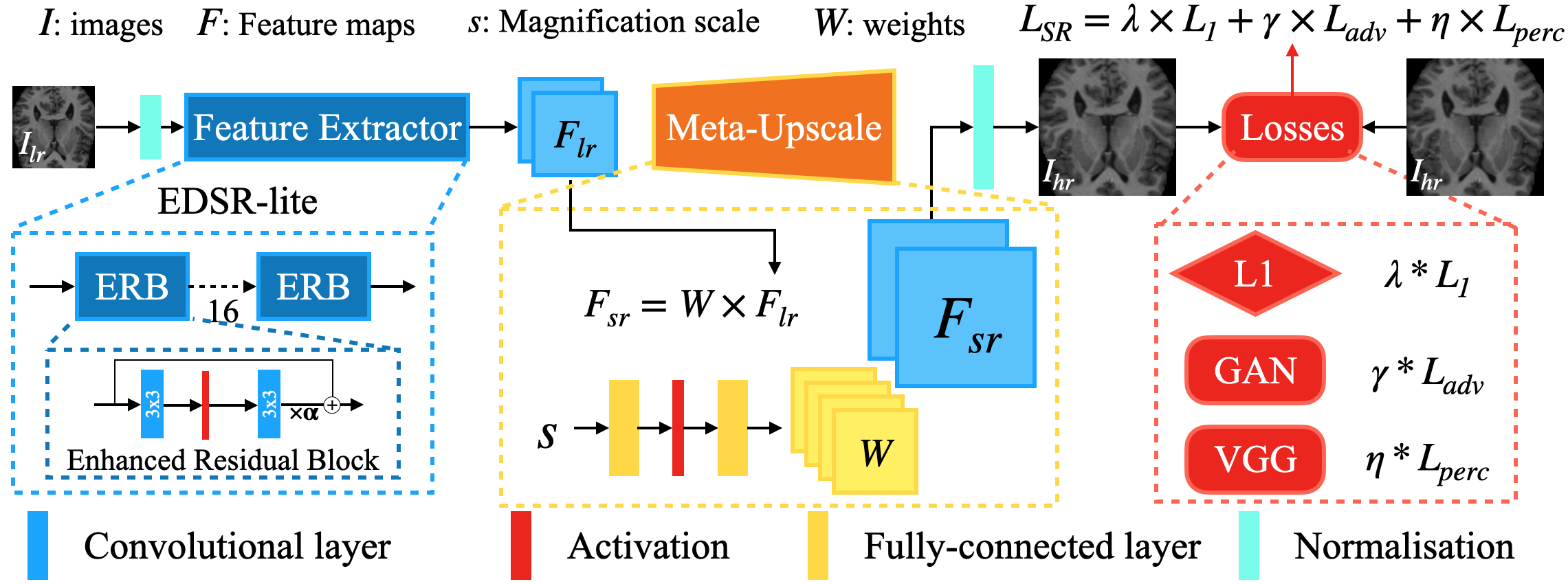}
	\caption{The proposed MIASSR consists of an EDSR-lite based low dimension feature extractor and a meta-upscale module. The feature extractor comprises 16 enhanced residual blocks, each of which includes two convolution layers and a non-linear activation layer. It extracts low dimension feature maps $F_{lr}$ of the input LR image $I_{lr}$. The meta-upscale module consists of two fully-connected layers and an activation layer. It predicts a group of weights from the input SR scale and achieves the feature map magnification with matrix multiplication. A super-resolved image $I_{sr}$ is then generated from the enlarged feature maps $F_{sr}$. The whole model is trained end-to-end with a combined loss function including L1 loss, adversarial loss and VGG based perceptual loss.}
	\label{fig:MIASSR_net}
\end{figure*}

\section{Related Work}
\subsection{Single Image Super-Resolution}
Traditionally, single image super-resolution methods increase the resolution of the input LR image using a variety of interpolation\cite{bicubic}, reconstruction\cite{wavelet}, neighbour embedding\cite{neighbour}, and sparse coding\cite{sparse}. However, these methods are incapable of simulating the non-linear transformation from LR image to high-resolution space and generate super-resolved images of poor quality. Recent developments in deep learning have led to dramatic improvements. Neural networks such as CNNs make it possible to learn this non-linear transformation from corresponded LR-HR image pairs. Dong et al. first pioneered SRCNN to solve SR problems\cite{SRCNN} with natural images, and achieved superior performance than traditional methods. Residual blocks\cite{ResNet} were then used to implement a much deeper network VDSR\cite{VDSR} and further improved the performance. These pre-upsampling methods first up-scaled the LR image with interpolation methods and then refined the primary results. However, processing feature maps in a high dimensional space has a high computational and memory cost, thus limiting the depth of networks. To avoid this inefficient calculation, post-upsampling architectures moved to process low-resolution feature maps and reconstruct SR images by up-sampling these LR features with transposed convolutions\cite{deconvolution} and sub-pixel layers\cite{SubPixel}. Meanwhile, various skip-connection strategies were introduced to stabilise the training of these deep networks, including residual connections\cite{VDSR}, dense connections\cite{SRDenseNet}, and residual dense connections\cite{RDN}. Despite the high performance on the peak signal-to-noise ratio (PSNR) and structural similarity (SSIM) of these deep networks, their outputs were not photo-realistic. Thus, SRGAN\cite{SRGAN} was proposed to overcome this gap and generate more photo-realistic images. Training with VGG\cite{VGG} based perceptual loss\cite{PerceptualLoss} and generative adversarial networks\cite{GAN}, SRGAN successfully generated perceptually realistic SR images which achieved comparable mean-opinion-score (MOS) with ground truth (GT) HR images. Following the same idea, ESRGAN\cite{ESRGAN} achieved superior performance by training a much deeper network with material recognition model based perceptual loss and relativistic discrimination\cite{RaGAN}. Additionally, methods with channel attention\cite{RCAN}, unsupervised learning\cite{ZSSR}, and back projection\cite{DBPN} were also proposed. Most of these methods were designed for specific magnification SR scales and are incapable of varying the up-sampling scale. Thus, MetaSR\cite{MetaSR} was proposed for scale-free SR tasks. It introduced a meta-upscale module that predicted a weight matrix and replaced the up-sample convolution layers (e.g., sub-pixel) with matrix multiplication. Instead of learning an up-sample transformation on a specific magnification scale, this new upscale module learns the relationship between up-sample transformations and scales of magnification, thus allowing a single model to super resolve images with arbitrary scales.

\subsection{Medical Image Super-Resolution}
The increasing interest in, and development of, CNN based SR algorithms has drastically improved their performance on medical image super-resolution tasks\cite{MedSRReview}. In contrast to natural image SISR tasks, medical image super-resolution is often pipelined by applications such as segmentation, classification and diagnosis, so it is required to preserve sensitive information and to enhance the structures of interest. Tackling the domain-specific SR problems, methods for medical image SR solved practical problems with techniques such as U-Net architecture\cite{UNet, UNetSR}, cycle GAN\cite{CycleGAN, CycleGANSR}, and 3D convolution\cite{3DSR}. Moreover, medical image SR has embedded benefits for other medical image analysis tasks such as lesion detection\cite{LFSR, MSGAN}, and segmentation\cite{SegSR}. However, most of the current methods have poor clinical applicability due to two limitations. First, they were mainly proposed for specific SR scales, which are not able to meet the range of magnification requirements in medical image analysis tasks and diagnosis. Second, modifying the methods designed for one medical image modality to new modalities and tasks is time-consuming and unreliable. Thus, in this paper, we propose a medical image arbitrary scale super-resolution (MIASSR) method, in which we seek to apply meta-learning to GANs to tackle arbitrary scale super-resolution tasks. Our main contributions are as follows:

\begin{itemlist}
\item We first introduce meta-learning to medical image SR tasks and propose the first scale-free SR method for medical images with SOTA performance. Notably, in this paper, we focus on SR tasks of up to $\times 4$ magnification to meet the most common requirements of medical image super-resolution in practice, but the proposed method can also work for a larger range of SR scales if proper training data is provided.
\item We show that, compared to an existing meta-learning based SR method\cite{MetaSR}, MIASSR has fewer parameters (only 26\%), but achieves comparable fidelity and perceptual quality, due to its use of adversarial learning and perceptual loss in the training process.
\item We successfully extend the proposed method to various medical scans of different purposes, such as single- and multi-modal MR images of the brain, cardiac MR images, and CT scans of COVID patients. With transfer learning, the pre-trained model can be applied to new modalities efficiently.
\end{itemlist}

\begin{figure*}[t]
	\centering
	\includegraphics[width=\textwidth]{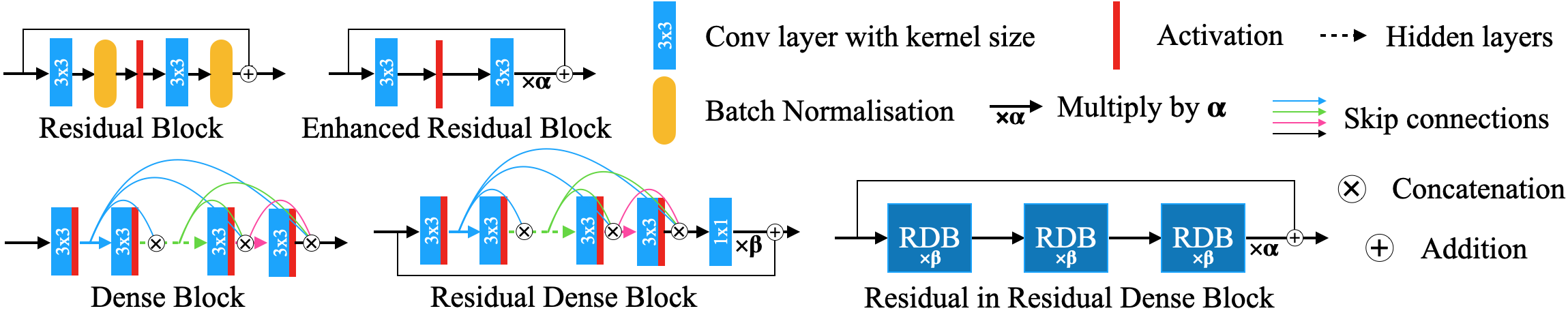}
	\caption{Residual and dense blocks. These blocks consist of convolution layers and activation layers in common but have various skip connections. They are the basic unit of SOTA SISR networks: Residual Block is used in SRResNet and SRGAN \cite{SRGAN}; Enhanced Residual Block is used in EDSR and MDSR \cite{EDSR}; Dense Block is used in SRDenseNet \cite{SRDenseNet}; Residual Dense Block is proposed in RDN \cite{RDN}; Residual in Residual Dense Block is proposed in ESRGAN \cite{ESRGAN}. Notice that residual scales $\alpha$ and $\beta$ are introduced to stablize the training of very deep neural networks (e.g., ESRGAN).}
	\label{fig:residual_blocks}
\end{figure*}

\section{Methods}\label{methods}
Single image super-resolution aims to restore a high-resolution image $I_{hr}$ from one low-resolution observation $I_{lr}$ of the same object. Generally in the real world, the LR image is modelled as\cite{SRReview18}:
\begin{equation}\label{eqt:degradation}
    I_{lr} = (I_{hr} \otimes \kappa) \downarrow_{s} + n,
\end{equation}
where $I_{hr} \otimes \kappa$ denotes the convolution between a blur kernel $\kappa$ and the HR image $I_{hr}$, $\downarrow_{s}$ is the down-sampling operation with scale $s$, and $n$ represents the noise. In SISR, we want to inverse the degradation mapping of equation (\ref{eqt:degradation}) to recover a super-resolved image $I_{sr}$ from $I_{lr}$\cite{SRReview}:
\begin{equation}
   I_{sr} = G(I_{lr}, s; \theta_{G}),
\end{equation}\label{eq:sr_basic}
where $G$ is a CNN based SR image generator and $\theta_{G}$ denotes its trainable weights. In each step of training, errors between the approximation $I_{sr}$ and the HR ground truth $I_{hr}$ are measured by a well-designed loss function $\mathcal{L}_{SR}$, and applied to the whole network using backpropagation to calculate gradients and update the weights $\theta_{G}$ \cite{SRReview}:
\begin{equation}
    \hat{\theta_{G}} = arg\ \underset{\theta_{G} } {\mathrm{min}}\mathcal{L}_{SR}(G(I_{lr}), I_{hr}).
\end{equation}\label{eqt:train_G}
The SR image generator in our approach consists of a feature extractor $\mathcal{F}$ which extracts the feature maps of the low-resolution image and a meta-upscale module $\mathcal{M}$ which up-samples the feature maps with arbitrary scales:
\begin{equation}
    I_{sr} = G(I_{lr}) = \mathcal{M}(\mathcal{F}(I_{lr}), s).
\end{equation}
Note that the input and output layers that normalise and convert the images to the feature domain, and vice versa in the generator, are ignored to simplify this equation. The rest of this section will introduce each component of the proposed MIASSR (Fig. \ref{fig:MIASSR_net}) in detail: firstly the feature extraction, then the meta-upscale module, and finally the losses for training.

\subsection{Feature Extraction}
We use an Enhanced Residual Block (ERB) $\mathcal{B}_{ERB}$ based feature extractor, namely EDSR-lite in MIASSR. This residual block (Fig. \ref{fig:residual_blocks}) is first proposed in EDSR\cite{EDSR}, which consists of two convolution layers $\mathcal{C}$, a non-linear activation layer $\sigma$, a residual connection, and no batch normalisation layer:

\begin{equation}\label{eqt:block}
\begin{split}
    & F_{out} = \mathcal{B}_{ERB}(F_{in}) \\
    & = F_{in} + \mathcal{C}^w_{1}(\sigma (\mathcal{C}^w_{2}(F_{in}; \phi_{\mathcal{C}_{2}})); \phi_{\mathcal{C}_{1}})\times \alpha, \\
    &{\phi_{\mathcal{C}_{1}}, \phi_{\mathcal{C}_{2}}} \in \theta_{G},
\end{split}
\end{equation}

where $F_{in}$ and $F_{out}$ are the input and output feature maps, $w$ presents the width of each convolutional layer, and $\left \{\phi_{\mathcal{C}_{1}}, \phi_{\mathcal{C}_{2}} \right \}$ are trainable parameters of the convolution layers. Padding is applied according to the kernel size in all convolution layers such that feature maps maintain their spatial dimensions. Thus, it becomes very convenient to build deeper neural networks by simply stacking several residual blocks, and low-dimension feature maps $F_{lr}$ can be extracted from the input LR image $I_{lr}$:

\begin{equation}\label{eqt:feature_extractor}
    F_{lr} = \mathcal{F}(I_{lr}) = \mathcal{B}^{d}(I_{lr}; \phi_{\mathcal{B}}), \phi_{\mathcal{B}} \in \theta_{G}.
\end{equation}
In our approach, we use a lite version of EDSR, which consists of 16 enhanced residual blocks (i.e., equation \ref{eqt:block}), and 64 convolution kernels with a size of $3 \times 3$ for each convolutional layer.

\subsection{Meta-Upscale Module}
To generate HR output from LR input, the low-dimension feature maps $F_{lr}$ extracted by $\mathcal{F}$ in equation (\ref{eqt:feature_extractor}) need to be upscaled:
\begin{equation}
    F_{sr} = \mathcal{U}_{s}(F_{lr}; \phi_{s}),
\end{equation}
where $F_{sr}$ is the super-resolved feature maps, $\mathcal{U}_{s}$ is an up-sampler, and $\phi_{s}$ is the group of parameters for the magnification option with scale $s$. In comparison with common single scale up-samplers such as sub-pixel\cite{SubPixel} which only learn one group of parameters for a specific SR scale $s$, the meta-upscale module\cite{MetaSR} $\mathcal{M}$ in our approach learns to predict a group of parameters for each SR scale. Particularly, one pixel $(i, j)$ of the super-resolved feature maps $F_{sr}$ is calculated as a weighted sum of all pixels in $F_{lr}$:
\begin{equation}
    F_{sr}(i, j) = \textbf{v}_{i, j} \times F_{lr}, 
\end{equation}
where $F_{lr}$ with original shape $(H_{in}, W_{in})$ is flattened to $(H_{in} \times W_{in}, 1)$ and $\textbf{v}_{i, j}$ is a vector with dimensions $(1, H_{in} \times W_{in})$. The parameters of $\textbf{v}_{i, j}$ are predicted by a weights prediction network $\mathcal{W}$ in $\mathcal{M}$ according to the scale $s$ and the pixel's location $(i, j)$:
\begin{equation}
     \textbf{v}_{i, j} = \mathcal{W}(\frac{i}{s}-\left \lfloor \frac{i}{s} \right \rfloor, \frac{j}{s}-\left \lfloor \frac{j}{s} \right \rfloor, \frac{1}{s}), (i, j)\in F_{sr}.
\end{equation}
Accordingly, all pixels in $F_{sr}$ can be achieved via matrix multiplication:
\begin{equation}
        F_{sr} = W_{s} \times F_{lr},
\end{equation}
Where $W_{s} = \mathcal{W}(s)$ denotes the magnification matrix consists of $\textbf{v}_{i, j}$ of all $(i, j) \in F_{sr}$ with scale $s$. Thus, the meta-upscale module is represented as:
\begin{equation}
    F_{sr} = \mathcal{M}(F_{lr}, s) = \mathcal{W}(s; \phi_{\mathcal{W}}) \times F_{lr}, \phi_{\mathcal{W}} \in \theta_{G},
\end{equation}
The weights prediction network $\mathcal{W}$ consists of only three-layers, including two fully-connected layers and a nonlinear activation layer.This meta-upscale module works with any scales, which is quite different with sub-pixel based up-samplers. Thus, it becomes possible to train an end-to-end model for SR tasks with arbitrary magnification scales.

\subsection{Loss Functions}
We use a combined loss of pixel-wise L1 loss, adversarial loss and perceptual loss as in equation (\ref{eqt:train_G}) to train our model.
\subsubsection{L1 loss}
SISR requires predicting the correct value of each pixel in the super-resolved images. Thus, pixel-wise errors are important for both evaluation and training of SR networks. In our work, we used the L1 loss, also called the mean-absolute-error (MAE) to train the model for good performance on PSNR and SSIM scores. It is defined as:
\begin{equation}
    \mathcal{L}_{1}(I_{sr}, I_{hr}) = \frac{1}{H*W}\sum_{(i,j)\in I}^{} \left \|  I_{hr}[i, j] - I_{sr}[i, j]\right \|,
\end{equation}
where $H$ and $W$ are the height and width of the images. L1 Loss is a typical loss function used to train SISR networks. However, as this loss leads to over-smoothing, it has limited ability in generating perceptually realistic textures in medical images, which are important for human beings for visual understanding. Therefore, the perceptual loss and adversarial loss are also introduced to the training of our method.

\subsubsection{VGG based perceptual loss}
The VGG based perceptual loss $\mathcal{L}_{perc}$ was first introduced by Johnson et al. \cite{PerceptualLoss}, and had been widely used in super-resolution tasks\cite{SRGAN, LFSR, MSGAN, ESRGAN}. It presents the mean-square-error(MSE) between the super-resolved images and the HR ground truth images in the feature domain:
\begin{equation}
    \mathcal{L}_{perc}(I_{sr}, I_{hr}) = \mathbb{E}(\left \| \mathcal{V}_{l}(I_{hr}) - \mathcal{V}_{l}(I_{sr})\right \|^2),
\end{equation}
where $\mathcal{V}$ is a pre-trained VGG19 model and $l$ denotes the feature maps of the specific layer of $\mathcal{V}$. Following the conclusion in ESRGAN\cite{ESRGAN}, we use the feature maps before activation of earlier layers to provide more textural information.

\subsubsection{Adversarial loss}
To generate more perceptually realistic images, we apply Wasserstein GAN based adversarial training in our method. In addition to the SR image generator $G$, a GAN consists of a discriminator $D$, which assists the training of $G$ by playing a game: the discriminator aims to distinguish generated fake images from real ground truth images, while the generator aims to fool the discriminator. The networks are trained jointly, so the discriminator should be able to recognise any image as real or fake correctly, while the generator should be able to produce as real as possible SR images. Thus, the basic adversarial loss function is defined as\cite{GAN}:
\begin{equation}
   \mathcal{L}_{GAN} = -\mathbb{E}_{I_{hr}}\left [ \log D(I_{hr}) \right ] - \mathbb{E}_{I_{lr}}\left [\log (1- D(G(I_{lr}))) \right ].
   \label{eqt:GAN}
\end{equation}
Whilst this basic version of adversarial loss, so-called vanilla GAN loss, has been successfully used in natural image\cite{SRGAN} and medical image\cite{LFSR} super-resolution tasks, it is susceptible to problems of training instability and mode collapse. Thus, Wasserstein GAN was proposed by Arjovsky et al.\cite{WGAN} to resolve these issues. Instead of the binary classification loss in vanilla GAN, the Wasserstein distance between the distribution of real and generated images was introduced as the adversarial loss:
\begin{equation}
    \mathcal{L}_{WGAN} = \mathbb{E}_{I_{lr}}\left [D(G(I_{lr})) \right ] -\mathbb{E}_{I_{hr}}\left [ D(I_{hr}) \right ]. 
    \label{eqt:WGAN}
\end{equation}
rOne important trick of Wasserstein GAN is to clip all the weights of the discriminator to a constant range $[-c, c]$, to meet the condition of derivable Wasserstein distance. However, with the clipping strategy, weights of discriminator tend to be either the minimum or maximum values. This causes the discriminator to behave like a binary network, and depresses the non-linear simulating abilities of GANs. Thus, the gradient penalty was proposed to replace the clipping operation\cite{WGANGP}. The new trick 
restricts the gradients of $D$ to not change rapidly, by adding a new term in the adversarial loss:
\begin{equation}
    \mathcal{L}_{adv} = \mathcal{L}_{WGAN} + \mathbb{E}_{I} \left [\left \| \bigtriangledown_{I}D(I) \right \|_{p} - 1 \right ]^2,
    \label{eqt:WGANGP}
\end{equation}
where $\left \| \right \|_{p}$ is the p-norm. 

In summary, the loss that we use to solve equation (\ref{eqt:train_G}) is defined as:
\begin{equation}\label{eqt:SRLoss}
    \mathcal{L}_{SR} = \lambda \times \mathcal{L}_{1} + \gamma \times \mathcal{L}_{adv} + \eta \times \mathcal{L}_{perc},
\end{equation}
where $\lambda$, $\gamma$, and $\eta$ are scale factors to balance each part of the loss function.

\begin{figure*}[t]
	\centering
	\includegraphics[width=\textwidth]{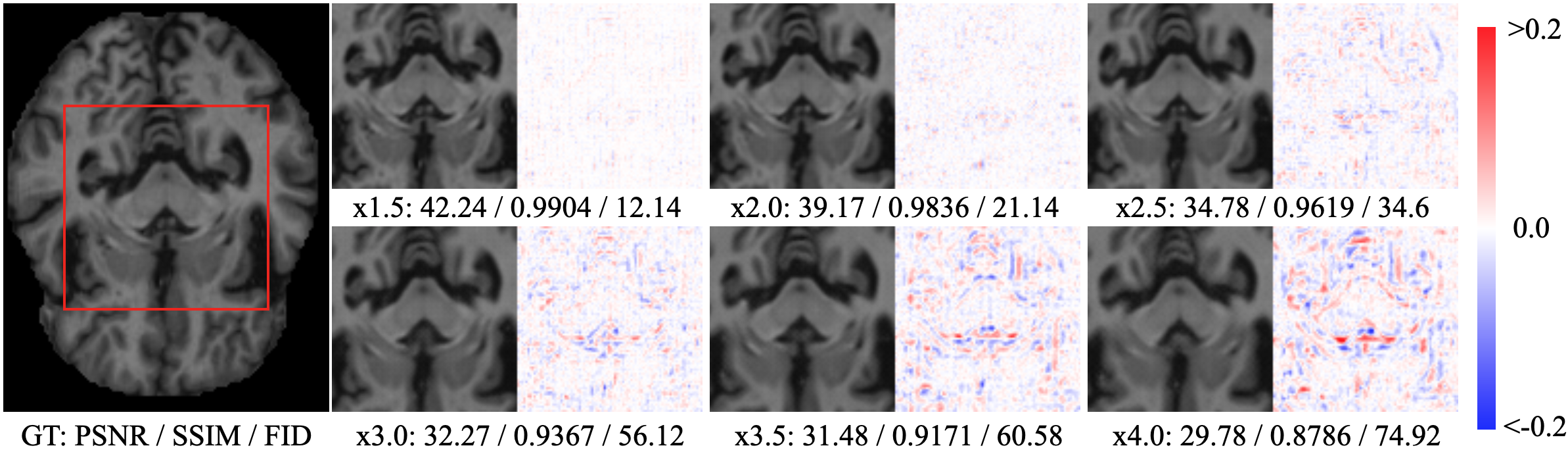}
	\caption{An example of super-resolved images with different scales of magnification by our proposed MIASSR. The slice is randomly selected from the test set of OASIS-brains. Here we only illustrate SR images with six scales $(1.5, 2.0, 2.5, 3.0, 3.5, 4.0)$, and the method could generate SR images with arbitrary scales in $(1, 4]$.} All images are converted to [0, 1]. Differences between SR images and ground truth images are rendered with the colour-bar and measured by PSNR, SSIM and FID. Higher PSNR and SSIM indicate better fidelity quality, while lower FID represents better perceptual quality.
	\label{fig:oasis_result}
\end{figure*}

\section{Experiments}
In our simulation experiments, the proposed method has been successfully applied to four different medical image datasets (Section \ref{sec:data}) in super-resolution tasks with arbitrary scales of magnification in $(1, 4]$.  HR ground truth (GT) images and corresponding LR images were generated from the original slices (Section \ref{sec:data}). To evaluate our method, metrics including PSNR, SSIM and Frechet Inception Distance (FID)\cite{FID} (Section \ref{sec:metrics}) were used to measure the differences between the super-resolved images and GT images in the test set. Afterwards, the mean performance over all scales in $(1, 4]$ was compared with  bicubic interpolation and seven SOTA SISR methods (Section \ref{sec:sota_methods}).

\subsection{Data and Pre-Processing}\label{sec:data}
LR-HR image pairs for training, validation and testing were generated from the original slices. HR images were achieved by removing the pure-black background margin of original slices, whilst LR images were generated by down-sampling the corresponding HR images and blurred with a $3 \times 3$ Gaussian kernel. We focused on the central regions of each slice as the pure-black background regions has little information content and including this area only slows the training process. In the experiments, a suitable margin size for each dataset was carefully chosen to ensure no non-zero values were removed.

\subsubsection{OASIS-brains}
The OASIS-brains\cite{OASIS} dataset consists of a cross-sectional collection of 416 subjects including individuals with early-stage Alzheimer's Disease (AD). For each subject, 3 or 4 individual T1-weighted MRI scans obtained within a single imaging session are included (Fig. \ref{fig:oasis_result}). For the single modality SR experiments, the brain-masked version of an atlas-registered gain filed-corrected image, namely T88-111, was used. Due to the limitations of computing resources, from the whole dataset, we randomly select 30 subjects for training, 3 subjects for validation and another 9 subjects for testing. Note that the validation dataset was only used for hyper-parameters searching. The original size of each subject was $[176 \times 208 \times 176]$, and only a central area of $[144 \times 180]$ were used. Our experiments were applied on the axial plane.

\subsubsection{BraTS}
The brain tumour segmentation dataset (BraTS)\cite{BraTSCite1, BraTSCite2, BraTSCite3} provids multi-modal MRI scans of 210 patients with glioblastoma (GBM$\slash$HGG) and the other 75 patients with lower grade glioma (LGG). Each BraTS multi-modal scan includes 4 MR modalities: native (T1), contrasted enhanced T1-weighted (T1ce), T2-weighted (T2), and T2 Fluid Attenuated Inversion Recovery (T2-FLAIR) volumes. We randomly selected 50 scans (35 HGG and 15 LGG) for training, and 10 scans (7 HGG and 3 LGG) for testing. The original image shape was $[240 \times 240 \times 155]$. Slices on axial plane were cropped to $[180 \times 170]$ to focus the training on brain area.

\subsubsection{ACDC}
The Automated Cardiac Diagnosis Challenge (ACDC)\cite{ACDC} dataset includes 1.5T and 3.0T MR scans of 150 subjects, including 30 health people and 120 patients with previous myocardial infarction, dilated cardiomyopathy, hypertrophic cardiomyopathy and abnormal right ventricle. The whole dataset was randomly assigned to 80 cases for training and 19 cases for testing. We applied MIASSR on the transverse plane, where the slices had various shapes from $[174 \times 208]$ to $[184 \times 288]$. To standardise the slices shapes, only the centre areas with a size of $[128 \times 128]$ were cropped and used for training and testing. 

\subsubsection{COVID-CT}
The COVID-CT dataset\cite{COVID} includes CT scans of 632 patients with COVID infections, from which images of 199 patients were used for training, and images of another 25 patients were used for testing. The original image shape is $[512 \times 512]$. After removing the background, only the centre area of $[412, 332]$ were used for either training or testing.

\subsection{Metrics}\label{sec:metrics}
Three objective image quality assessment methods were used to measure both the fidelity and perceptual quality of the generated SR images in our experiments. First, we use the peak signal-to-noise ratio (PSNR), which is defined as\cite{SRReview}:
\begin{equation}\label{eqt:psnr}
    PSNR(I_{sr}, I_{hr}) = 10 \cdot log_{10}(\frac{L^2}{\frac{1}{N}\sum_{i=1}^{N}(I_{sr}(i)-I_{hr}(i))^2}),
\end{equation}
where $L$ denotes the maximum pixel ($L = 1.0$ in our case), and N is the number of all pixels in $I_{sr}$ and $I_{hr}$. PSNR relates to the pixel-level mean squared error, and is the most widely used evaluation criteria for SR models\cite{SRReview, SRReview18, MedSRReview}. Meanwhile, another popular metric, structural similarity (SSIM), is also used. It is defined as\cite{SRReview18}:
\begin{equation}\label{eqt:ssim}
    SSIM(x, y) = \frac{2\mu_{x}\mu_{y} + \kappa_1}{\mu_x^2 + \mu_y^2 + \kappa_1} \cdot \frac{\sigma_{xy}+\kappa_2}{\sigma_x^2 + \sigma_y^2 + \kappa_2}, 
\end{equation}
where $x, y$ denote two images, $\mu$ and $\sigma^2$ are the mean and variance, $\sigma_{xy}$ is the covariance between $x$ and $y$, and $\kappa_1, \kappa_2$ are constant relaxation terms. SSIM assumes that the human visual system is highly adapted to extract image structures, and measures the structural similarity between images based on the comparisons of luminance, contrast, and structures.

However, both metrics are limited to measure the fidelity quality, but cannot evaluate perceptual quality. Over-smoothed images were reported\cite{SRGAN, MSGAN} to achieve higher PSNR and SSIM scores than texture-rich images, but they might be less perceptually realistic. Thus, we also calculated the Frechet Inception Distance (FID)\cite{FID}, which is widely used to evaluate the perceptual quality of generated images by GANs. It measures the difference of high-level global features of generated SR images and GT images, by calculating the distance between the distributions of both groups of images in the latent space of a pre-trained image classification model Inception-V3\cite{InceptionV3}. Note that higher scores of PSNR and SSIM represent better fidelity quality, while lower FID indicates more perceptually realistic images have been generated.

\begin{figure*}[t]
	\centering
	\includegraphics[width=\textwidth]{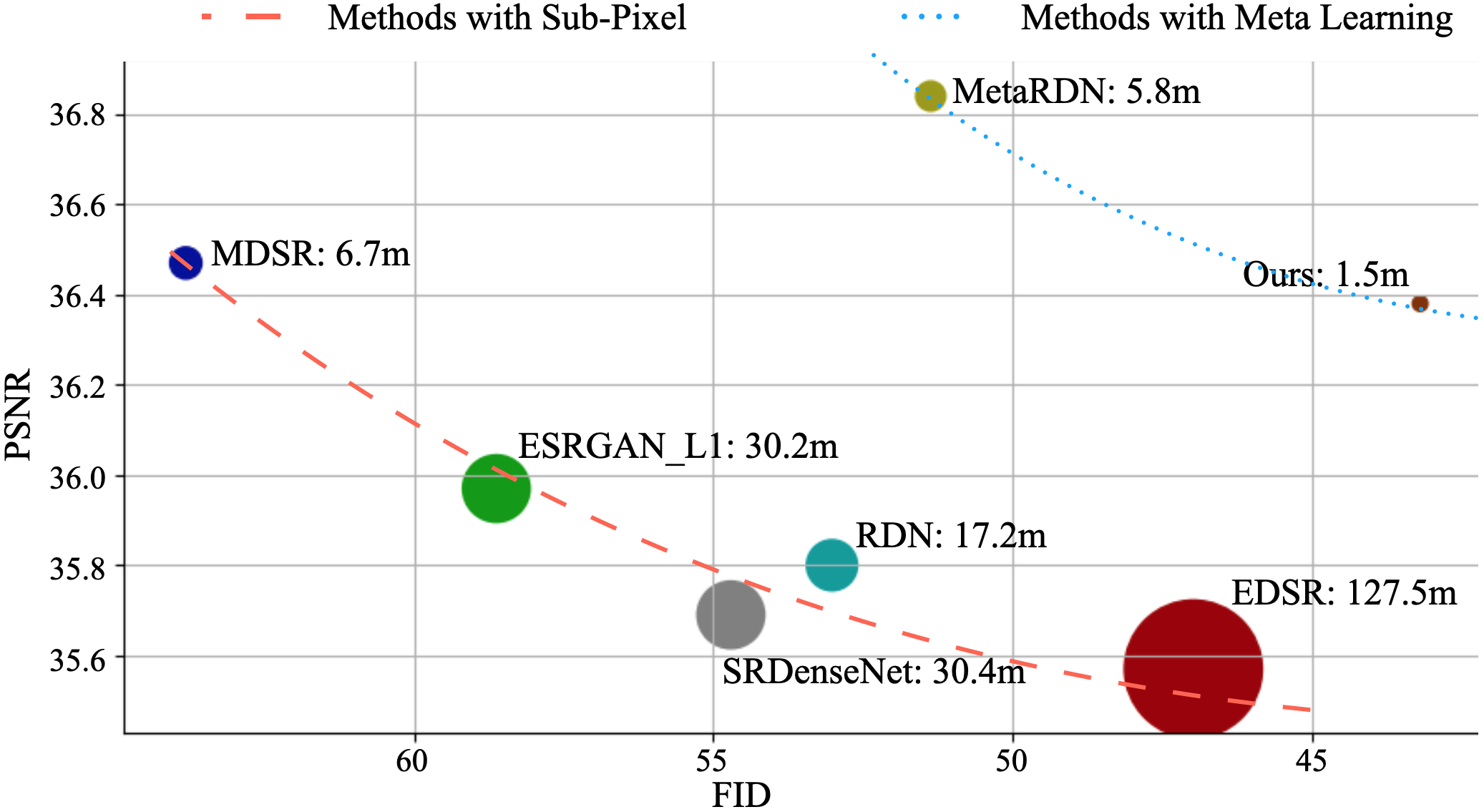}
	\caption{Performance and model size of our method compared to SOTA methods. Mean PSNR and mean FID on SR scales between $(1, 4]$ are used to evaluate both the fidelity and perceptual quality of generated images. The bubble size denotes the number of parameters of each model. SRGAN is not plotted because it has much worse performance ($mean PSNR = 28.15, mean FID = 128.21$) compared to all other methods. Higher PSNR scores denote better fidelity quality, and lower FID represents better perceptual quality. Our proposed method has achieved the best FID and the third-best PSNR with the smallest model size. Perception-distortion trade-off\cite{TradeOff} is reflected in both groups: methods with meta-learning, and methods with sub-pixel. High fidelity and perceptual quality are at odds, and impossible to be improved simultaneously.}
	\label{fig:num_paras}
\end{figure*}

\begin{table*}[t]
\tbl{Comparing our method with SOTA methods in SR tasks with arbitrary scales in $(1, 4]$ on the OASIS-brains dataset. Higher PSNR and SSIM denote better fidelity quality, while lower FID means better perceptual quality. The best performance is in bold. The unit of the number of parameters of each model is million. Along with the smallest model size out of all CNN-based methods, our proposed method achieves the best FID and comparable PSNR/SSIM.\label{tab:sota}}
{\begin{tabular}{@{}l|rrrrrrrrr@{}}
\Hline
  & BiCubic & SRGAN \cite{SRGAN} & EDSR \cite{EDSR}   & ESRGAN(L1) \cite{ESRGAN} & MDSR \cite{EDSR}  & RDN  \cite{RDN}  & SRDenseNet \cite{SRDenseNet} & MetaRDN  \cite{MetaSR}       & Ours           \\ \hline
PSNR   & 31.32   & 28.79  & 35.71  & 36.11      & 36.63  & 35.95  & 35.83      & \textbf{36.84}  & 36.46          \\
SSIM   & 0.8600  & 0.6380 & 0.9541 & 0.9568     & 0.9595 & 0.9574 & 0.9548     & \textbf{0.9627} & 0.9576         \\
FID    & 144.9   & 117.7  & 44.00  & 54.36      & 58.95  & 49.37  & 50.90      & 51.36           & \textbf{39.85} \\
Params & $-$     & 4.5M   & 127.5M & 30.2M      & 6.7M   & 17.2M  & 30.4M      & 5.8M            & \textbf{1.5M}  \\
\Hline
\end{tabular}}
\end{table*}

\subsection{Implementation Details}
We use PyTorch to implement our method, NiBabel to load medical data, and OpenCV-python for image resize and blur operations. All experiments were performed on an Nvidia Quadro RTX 8000 GPU. Following the instructions of our open-source implementation,\footnote{https://github.com/GinZhu/MIASSR} researchers could reproduce our experiments and test new experiments easily. The following hyper-parameters and details of training tricks were also released as config files.

The generator consists of $d=16$ enhanced residual blocks, in which each convolution layer has $w=64$ feature maps. A residual scale $\alpha=1.0$ was used for the residual connections in ERB. The discriminator is similar as in DCGAN\cite{DCGAN} and SRGAN\cite{SRGAN}. It consists of 7 down-sample blocks, each of which has one convolution layer with stride $=1$ for feature expanding and one convolution layer with stride $=2$ for down-sampling. No batch normalisation layers were used in either the generator or the discriminator, while leaky-ReLU with negative-slope $=0.2$ was chosen as the non-linear activation function. In the experiments without transfer learning, both networks were initialized with Kaiming-uniform\cite{HeInit}.

During training, LR and HR images were randomly cropped into sample patches. The original path size was set to $H_{p}, W_{p} = (96, 96)$ for HR patches, but either the size of LR patches or the size of the HR patches were adjusted to match the magnification scale $s$:
\begin{equation}
    \begin{split}
        & H_{lr}, W_{lr} = \left \lfloor \frac{H_{p}}{s} \right \rfloor, \left \lfloor \frac{W_{p}}{s} \right \rfloor;\\
        & H_{hr}, W_{hr} = \left \lfloor sH_{lr} \right \rfloor, \left \lfloor sW_{lr} \right \rfloor.
    \end{split}
\end{equation}

For each training step, a batch of 16 random patches, with the same SR scale was fed to the model. Initially, in pre-training, the generator of MIASSR was trained with only $\mathcal{L}_{1}$ for $1\times 10^{5}$ steps, because this `warm-up' has been found to stabilize the training of GANs\cite{ESRGAN}. Then we train both generator and discriminator with $\lambda =1$, $\gamma=0.001$ and $\eta = 0.006$ for $1 \times 10^{5}$ steps. Adam optimiser\cite{Adam}, with an initial learning rate $l_{r}=0.0001$, momentum $m=0.9$, and betas $b=(0.9, 0.999)$ was used for backpropagation. Every $5 \times 10^{4}$ steps, the learning rate was halved. To avoid a gradient explosion, losses above $1\times 10^8$ were discarded.

All the above hyper-parameters were chosen based on the validation performance in our experiments on the OASIS-brains dataset. In the transfer learning experiments of ACDC and COVID-CT datasets, the model which had been pre-trained on OASIS-brain was fine-tuned for $1 \times 10^4$ steps with $\mathcal{L_{SR}}$. Meanwhile, to make MIASSR work with multi-modal scans in the experiments with the BraTS dataset, the single-channel input and output layers were modified to 4-channel. Particularly, the four modalities of BraTS, T1ce, T1, T2 and Flair, were stacked during both training and testing, and the loss function $\mathcal{L}_{SR}$ was first calculated on each modality independently and then averaged.

\subsection{Comparison with SOTA Methods}\label{sec:sota_methods}
We compared the proposed method with the bicubic interpolation and 7 SOTA SISR methods: SRGAN\cite{SRGAN}, EDSR\cite{EDSR}, SRDenseNet\cite{SRDenseNet}, RDN\cite{RDN}, MDSR\cite{EDSR}, ESRGAN-L1\cite{ESRGAN}, and MetaRDN\cite{MetaSR}. All these methods were designed for natural images, which have  much bigger dimensions than the medical images we used. To make them work well with our experiments, we re-trained the models with the medical image datasets with smaller patches. All models were trained with the same steps and learning rate decay policy, without model embedding and data augmentation for a fair comparison. We used the original loss functions to train most of the models, but the material recognition\cite{ESRGANPercptualLoss} based perceptual loss in ESRGAN hindered training with medical images, so we used the L1-loss based ESRGAN (so-called ESRGAN-L1) for our comparison.

All these SISR methods (except MetaRDN) were designed only for specific integer magnification scales ($\times 2$, $\times 3$, and $\times 4$). To evaluate their performance on SR tasks with arbitrary scales, we used an up-and-down strategy, consisting of two steps: in the up-sampling step, the well-trained model with the ceiling scale $\left \lceil s \right \rceil$ is used to infer an over-magnified SR image; then, in the down-sampling step, the over super-resolved image was resized correspondingly using the bicubic interpolation.

\begin{table*}[t]
\tbl{Comparing MIASSR with EDSR, MetaRDN, and bicubic interpolation on multi-modal brain images (BraTS), cardiac MR scans (ACDC), and chest CT images of COVID patients (COVID-CT). Higher PSNR represents better fidelity quality of SR images, while lower FID denotes better perceptual quality. The best performance is in bold. Notice that MIASSR requires only one fifth training steps of EDSR and MetaRDN with transfer learning. Moreover, unlike EDSR and MetaRDN which only work with a single modality, it works with multi-modality data in a single model and further reduces the cost of extending to new SR tasks.\label{tab:brats_acdc_covid}}
{\begin{tabular}{@{}l|rrrrrr@{}}
\Hline
PSNR / FID   & BraTS-T1       & BraTS-T1ce    & BraTS-T2      & BraTS-Flair   & ACDC          & COVID-CT      \\ \hline
BiCubic      & 32.83 / 151.6  & 33.13 / 139.7 & 29.87 / 125.1 & 31.52 / 145.8 & 27.42 / 267.7 & 38.62 / 141.0 \\
EDSR  \cite{EDSR}       & 36.54 / 76.42  & 36.34 / 74.18 & 33.69 / 51.93 & 35.27 / 79.32 & 30.82 / 179.3 & \textbf{46.75} / 39.53 \\
MetaRDN \cite{MetaSR}     & \textbf{37.27} / 78.16  & \textbf{37.22} / 72.93 & \textbf{34.76} / 53.55 & \textbf{36.05} / 78.88 & \textbf{31.27} / 154.7 & 43.27 / 38.96 \\
MIASSR(ours) & 36.89  / \textbf{62.28} & 36.86 / \textbf{61.06} & 34.28 / \textbf{46.74} & 35.91 / \textbf{68.45} & 30.94 / \textbf{153.5} & 43.21 / \textbf{37.66}  \\
\Hline
\end{tabular}}
	
\end{table*}

\section{Results and Discussion}
In this section, we first present the results of MIASSR with four datasets, and compare the performance with SOTA SR methods. Then, we systemically analyse the influence of each component of the model. The ablation study is important to understand the how the hyper-parameters and model architecture affect the final SR performance, as in other image processing tasks. \cite{IJNSImagePairs}

\subsection{MIASSR Performance}
First of all, in the experiments with the OASIS-brains dataset (Fig. \ref{fig:oasis_result}), MIASSR was compared with SOTA methods with arbitrary SR scales. On average, it achieved the third-best mean PSNR and SSIM scores, and the best FID with the fewest parameters (Fig. \ref{fig:num_paras}, and Table. \ref{tab:sota}). Particularly, with all SR scales, MIASSR generated images with comparable fidelity and perceptual quality with SOTA methods (Fig. \ref{fig:sota}). Considering the balance between cost and performance, the mean PSNR score and FID were plotted with the number of parameters of each model. Particularly for SISR methods which only supported one SR scale (i.e., EDSR, RDN, SRDenseNet and ESRGAN), the model size denoted the number of all parameters in three models $\times 2$, $\times 3$, and $\times 4$, because they were all required in inference. Methods designed for multi-scale (i.e., MDSR) and arbitrary scales (i.e., MetaRDN and Ours) greatly decreased the effective model size by reducing the required models to one.

\begin{figurehere}
	\centering
	\includegraphics[width=3.2in]{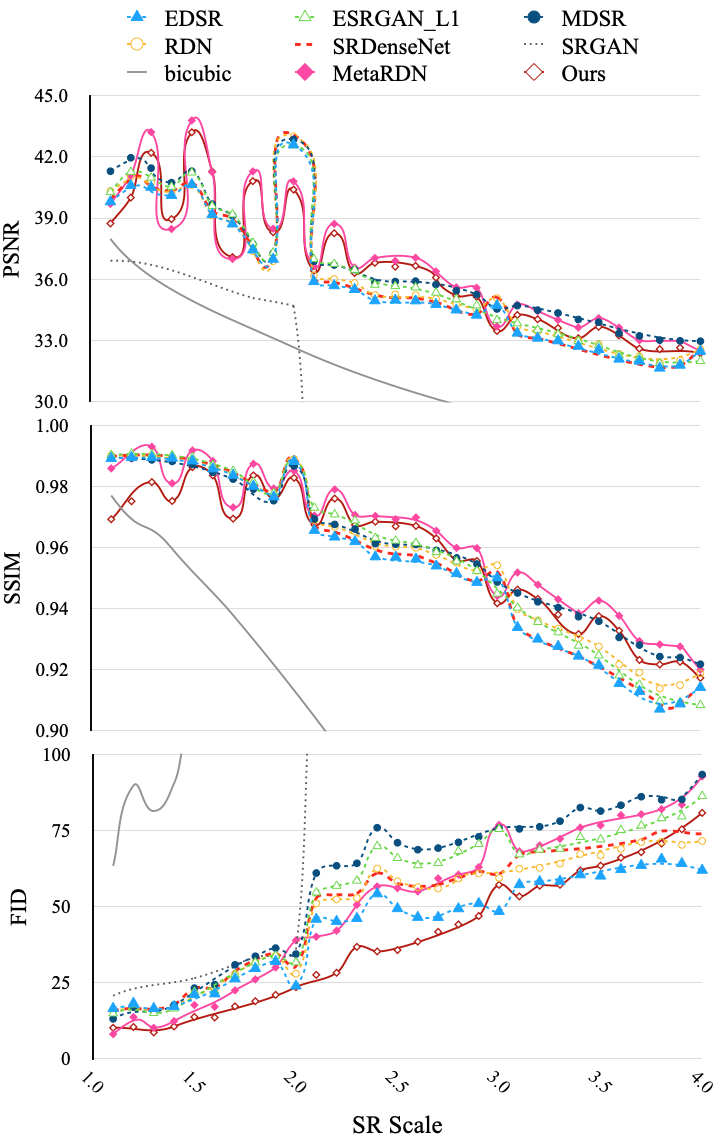}
	\caption{Comparing our proposed method with bicubic interpolation and SOTA methods in SR tasks with arbitrary scales in $(1, 4]$. Higher PSNR and SSIM denote better fidelity quality, while lower FID represents better perceptual quality. Results of bicubic interpolation and SRGAN are not fully plotted because of their poor performance. Our proposed method has achieved comparable performance with SOTA methods with all SR scales.}
	\label{fig:sota}
\end{figurehere}

 Notably, MIASSR had the fewest parameters which were only $26\%$ of the existing meta-learning based method (i.e., MetaRDN) and fewer than $1\%$ of EDSR.
Interestingly, multi-task methods (i.e., MDSR, MetaRDN, our MIASSR) achieved much better PSNR scores than single-scale SR methods, although they were learning more challenging transformations. Instead of approximating one mapping with a certain SR scale, their parameters were shared in approximations of mappings with various SR scales. However, the parameter sharing between tasks and the diversity of LR-HR image pairs with different magnification scales helped the models converged better in the training process. Interestingly, if we divided all the methods into two groups: methods with meta-learning and methods without meta-learning, the perception-distortion trade-off\cite{TradeOff} was clearly observed in each group. We found that the stem of workflow determined the overall performance, while the details of architectures only affect the balance between fidelity and perceptual quality.

The proposed method has also performed well with various medical image modalities (Table. \ref{tab:brats_acdc_covid}). Transfer learning, which helps to decrease the training cost on medical data processing \cite{TransferLearning}, also helped to modify the OASIS-brains pre-trained model to new single-modality medical image datasets efficiently and effectively. Compared to the bicubic interpolation method, the proposed MIASSR significantly improves performance in experiments with ACDC and COVID-CT. Comparing with EDSR and MetaRDN, MIASSR generated images of comparable fidelity and better perceptual quality, with only one fifth of the training steps. Indeed, transfer learning reduced the required training steps from $1 \times 10^5$ to $2 \times 10^4$. Besides, MIASSR is straightforwardly extended to multi-modality images: by simply modifying the input and output layers, it successfully worked for the cross-modality SR task of the BraTS dataset. Comparing with SOTA methods that work for a single modality, it achieved comparable performance on all four modalities (T1, T2, T1ce and Flair). The additional visualisation results of ACDC, COVID-CT and BraTS are attached in the Appendix.

In summary, the proposed method has good clinical applicability. In our experiments, it has been successfully applied to various medical image super-resolution tasks, with different situations of modalities and diseases. It works with the brain and cardiac MR images (i.e., OASIS-brains, ACDC), chest CT images (i.e., COVID-CT), and cross-modality scans (BraTS). Comparing with SOTA methods, MIASSR can generate SR images with comparable fidelity and better perceptual quality with much smaller model size. Additionally, with transfer learning, MIASSR can be extended to new datasets effectively and efficiently.

\begin{figurehere}
	\centering
	\includegraphics[width=3.2in]{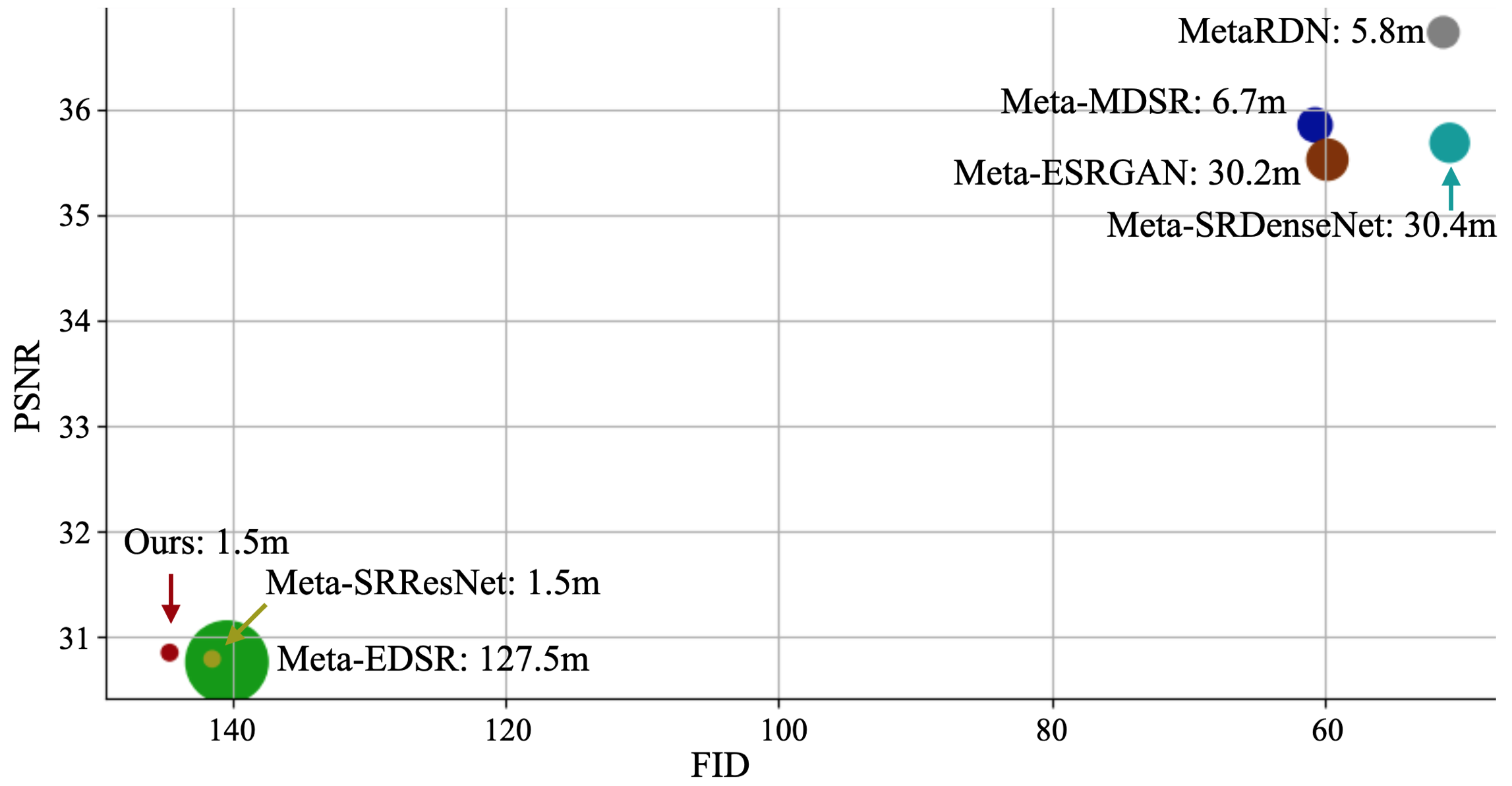}
	\caption{Comparison of SR image generators in MIASSR (training with $\mathcal{L}_{1}$). Higher PSNR indicates better fidelity quality, while lower FID represents more perceptually realistic SR images. Bubble size denotes the number of parameters. The basic block design, such as skip connections, rarely affect the final performance, but the depth of networks impact the performance a lot. Deeper networks (MetaRDN, Meta-MDSR, Meta-ESRGAN, and Meta-SRDenseNet) perform better.}
	\label{fig:generators}
\end{figurehere}

\subsection{Generator Architectures}\label{sec:ds_generators}
Referring to equation (\ref{eqt:block}) and (\ref{eqt:feature_extractor}), the SR image generator is influenced by three factors: the block structure $\mathcal{B}$, the number of blocks (namely the depth $d$), and the number of convolution kernels of each layer (namely the width $w$). Broadly, the width and depth of the network determine the size of the network, while the block structure represents the connection of these layers. They resolve the capability of feature extraction jointly. To understand how each factor influences the final performance, we have designed an ablation study of generator architectures.

\begin{figurehere}
	\centering
	\includegraphics[width=3.2in]{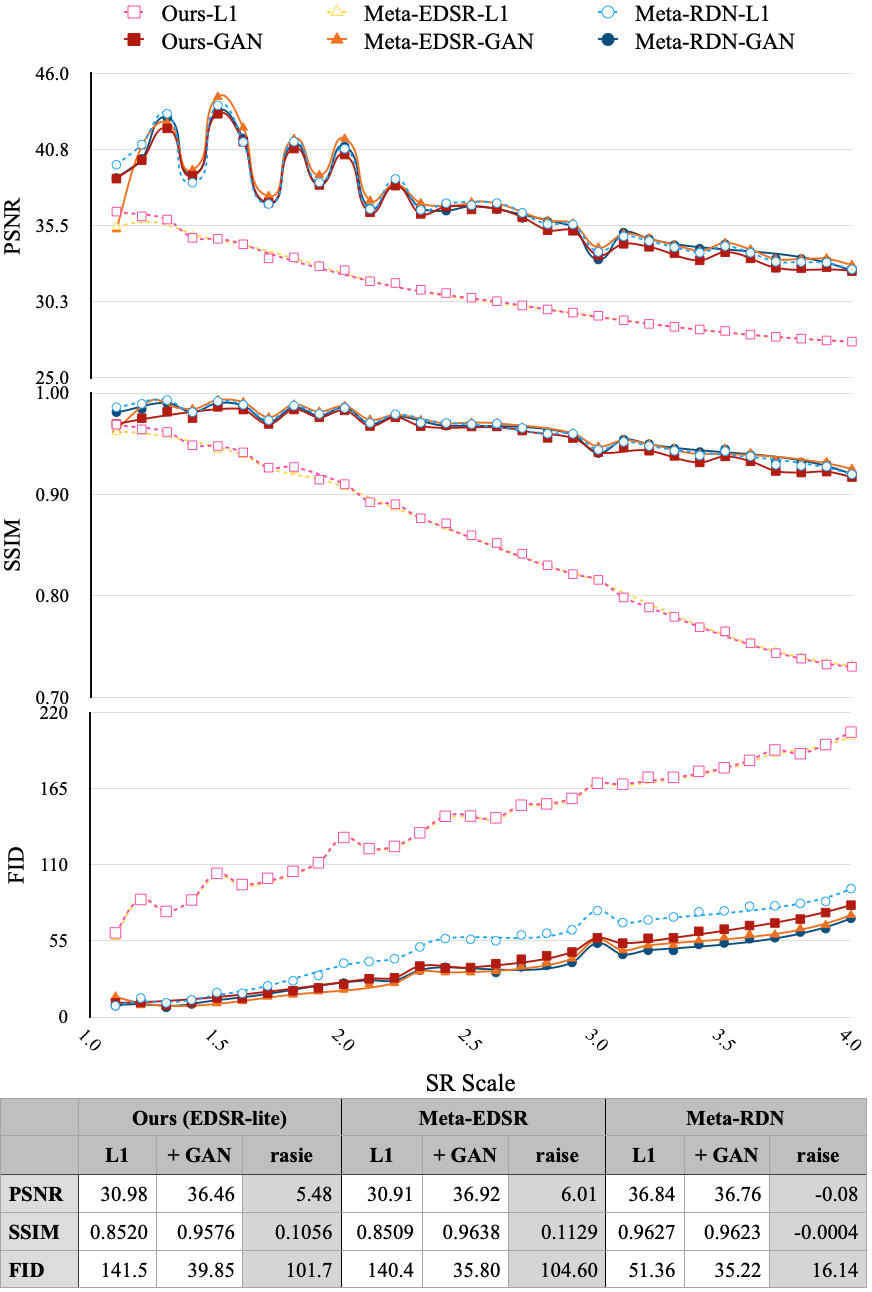}
	\caption{The sensitivity analysis of SR image generators to GAN based adversarial learning. Three networks, EDSR-lite (ours), EDSR and RDN are trained with $\mathcal{L}_{1}$ only, and with additional $\mathcal{L}_{adv}$ and $\mathcal{L}_{perc}$, and compared. Low FID represents good perceptual quality of generated images, while high PSNR and SSIM indicate good fidelity quality. Adversarial learning has significantly improved the performance of EDSR based methods.}
	\label{fig:l1_gan_upgrade}
\end{figurehere}

First, we have compared a wide range of architectures for LR feature extraction in MIASSR. Six SOTA networks, which have been widely used in computer vision and achieved high performance on SISR tasks, were tested. These networks,
SRResNet, EDSR, MDSR, SRDenseNet, RDN and ESRGAN, were built with various residual and dense blocks (Fig.\ref{fig:residual_blocks}) and behaved differently on simulating the LR-to-HR transformation. To match MIASSR, we replaced their scale specific up-samplers with the meta-upscale module, as in MetaSR\cite{MetaSR}, and kept the original settings (i.e. depth and width). To compare the performance of these generators, we trained them with the L1 loss $\mathcal{L}_{1}$ for $1 \times 10^5$ steps with the same training settings. Both fidelity and visual qualities of the generated SR images were evaluated (Fig. \ref{fig:generators}).

When training with only fidelity loss $\mathcal{L}_{1}$, we found it very interesting that these variations could be divided into two groups. Methods with more layers always performed better, because the deeper structure provided stronger ability in simulating non-linear transformations. Other structures, such as skip connections in dense blocks and the width of models, made no big differences. However, suitable structures for minimising pixel-wise errors might not fit the needs of generating perceptually realistic textures. Thus, we further compared RDN, EDSR and EDSR-lite with adversarial learning (Fig.\ref{fig:l1_gan_upgrade}). Although MetaRDN achieved the best performance with L1 loss, it was not improved with further adversarial learning. Oppositely, the other two were distinctly benefited by the adversarial and perceptual loss. However, comparing the original EDSR with the EDSR-lite, extra feature maps of each convolution layer only slightly improved the performance, although nearly one hundred times more parameters were used.

\begin{tablehere}
\tbl{Effects of the width of the network $w$, which represents the number of convolution kernels in each layer. Increasing width leads to additional parameters quadratically. Higher PSNR, SSIM and lower FID denote better performance. Bold texts represent the best performance, and the grey column denotes the hyper-parameter of our method. We finally choose $w = 64$ because it has achieved the best PSNR and FID.\label{tab:n_feats}}
{\begin{tabular}{@{}l|rrrr>{\columncolor[gray]{0.8}}rr@{}}
\Hline
$w$ =  & 4               & 8      & 16     & 32     & 64             & 128             \\ \hline
PSNR   & 34.58           & 34.83  & 35.22  & 35.89  & \textbf{36.46} & 36.44           \\
SSIM   & 0.9346          & 0.9399 & 0.9469 & 0.9542 & 0.9576         & \textbf{0.9592} \\
FID    & 84.17           & 73.49  & 58.23  & 49.57  & \textbf{39.85} & 43.84           \\
Params & \textbf{0.016M} & 0.041M & 0.12M  & 0.42M  & 1.5M           & 5.8M \\
\Hline
\end{tabular}}
\end{tablehere}

Additionally, we investigated the influence of the width and depth of the network over the final SR performance. Technically, deeper and wider networks should have more capability of approximation, which would result in better SR performance. However, more trainable parameters also led to more challenging optimisation of the equation (\ref{eqt:train_G}) and over-fitting. Meanwhile, the cost and performance balance should also be considered, because the number of parameters grows linearly with the depth and quadratically with the width. In our experiments of the width of the network, the model that consisted of 64 convolution kernels in each layer achieved the best PSNR and FID (Table. \ref{tab:n_feats}). Wider networks did not improve the performance (e.g. $w=128$), and the optimisation in networks training even failed to converge when $w = 256$. Similarly in the experiments of the depth of the network, additional residual blocks over 16 rarely helped (Table. \ref{tab:n_blocks}).

\begin{tablehere}
\tbl{Effects of the depth of the network $d$, which represents the number of residual blocks in the network. The unit of the number of parameters is million. Higher PSNR, SSIM and lower FID denote better performance. Bold texts represent the best performance, and the grey column denotes the hyper-parameter of our method. We finally choose $d = 16$, because extra blocks rarely improve the performance, but lead to more parameters linearly.\label{tab:n_blocks}}
{\begin{tabular}{@{}l|rrr>{\columncolor[gray]{0.8}}rrr@{}}
\Hline
$d$ =      & 2      & 4      & 8      &  16                          & 32     & 64     \\ \hline
PSNR       & 35.09  & 35.48  & 36.03  & 36.46  & \textbf{36.73}  & 36.46  \\
SSIM       & 0.9438 & 0.9490 & 0.9549 & 0.9576 & \textbf{0.9622} & 0.9621 \\
FID        & 62.47  & 54.44  & 49.61  & 39.85  & 37.46  & \textbf{37.39}  \\
Params & \textbf{0.48M}  & 0.63M  & 0.93M  & 1.5M   & 2.7M   & 5.1M \\
\Hline
\end{tabular}}
\end{tablehere}

Network architectures, such as skip connections, depth and width have significant impacts on the final performance of medical image SR tasks. First of all, extra skip connections is a double-edged sword. These connections in dense blocks such as RDB and RRDB add more pathways in the networks. With these highways of loss information, gradients could be passed more efficiently and effectively to each layer during backpropagation. Thus, the model (e.g. RDN) could easily achieve a very good performance, especially for minimising simple and clear errors such as L1 loss. However, the structure of the dense connections also made the model liable to getting stuck in certain points, and made the model insensitive to uncertain losses such as GANs. As a result, smaller models with fewer connections, such as EDSR-lite, could be comparable with them. Second, wider models are not necessary for medical images. Both RDN and original EDSR had many more feature maps than EDSR-lite in each convolutional layer, which made them more powerful in simulating and extracting features of nature images in SR tasks. However, for medical images with limited size and 
relatively lower contrast information, too many feature maps tended to be overqualified. In summary, we decided to use EDSR-lite, consisting of 16 residual blocks and 64 convolution kernels in each layer, because it had the fewest parameters and achieved equal performance with bigger models.

\begin{tablehere}
\tbl{Effects of perceptual loss. In this experiment, we set $\lambda = 1$ and $\gamma = 0.001$, and test different values of $\eta$. Particularly $\eta = 0$ means no perceptual loss, while $\eta = \infty$ means only perceptual loss is used for training. Higher PSNR, SSIM and lower FID denote better performance. Bold texts represent the best performance, and the grey column denotes the hyper-parameter of our method. \label{tab:perc_loss}}
{\begin{tabular}{@{}l|r>{\columncolor[gray]{0.8}}rrrrr@{}}
\Hline
$\eta = $ & 0      & 0.006  & 0.01   & 0.1    & 1      & $\infty$ \\ \hline
PSNR      & 36.52  & 36.46  & \textbf{36.64}  & 36.34  & 36.11  & 34.94    \\
SSIM      & \textbf{0.9611} & 0.9576 & 0.9607 & 0.9583 & 0.9554 & 0.9532   \\
FID       & 47.87  & \textbf{39.85}  & 40.61  & 43.59  & 45.79  & 46.61   \\
\Hline
\end{tabular}}
\end{tablehere}

\begin{tablehere}
\tbl{Effects of adversarial loss. In this experiment, we set $\lambda = 1$ and $\eta = 0.006$, and test different values of $\gamma$. Particularly $\gamma = 0$ means no adversarial loss, while $\gamma = \infty$ means only adversarial loss is used for training. Higher PSNR, SSIM and lower FID denote better performance. Bold texts represent the best performance, and the grey column denotes the hyper-parameter of our method. \label{tab:adv_loss}}
{\begin{tabular}{@{}l|r>{\columncolor[gray]{0.8}}rrrrr@{}}
\Hline
$\gamma = $ & 0      & 0.001  & 0.01   & 0.1    & 1      & $\infty$ \\ \hline
PSNR        & \textbf{36.60}  & 36.46  & 35.74  & 36.26  & 30.92  & 30.52    \\
SSIM        & \textbf{0.9592} & 0.9576 & 0.9555 & 0.9540 & 0.8798 & 0.8768   \\
FID         & 41.94  & 39.85  & \textbf{37.47}  & 44.17  & 114.9  & 103.1    \\
\Hline
\end{tabular}}
\end{tablehere}

\subsection{Loss Functions}
Referring to equation (\ref{eqt:train_G}) and (\ref{eqt:SRLoss}), the joint loss function $\mathcal{L}_{SR}$ plays an important role in the training of MIASSR. In the three components, $\mathcal{L}_{1}$ represents the pixel-wise errors, while $\mathcal{L}_{perc}$ and $\mathcal{L}_{adv}$ denote the visual dissimilarity of the entire images. Particularly, the perceptual loss $\mathcal{L}_{perc}$ considers the general visual features of images, because it is based on a well-trained VGG network with big plenty of normal images. In contrast, the adversarial loss $\mathcal{L}_{adv}$ is trained with the model, so it focuses on the inner features of the training dataset more. To achieve the best performance and to understand each component well, we have tested various weights of each loss. We first set $\lambda = 1$, then tested different values of $\gamma$ and $\eta$. Regarding the perceptual loss, we have tested $\eta = (0, 0.006, 0.01, 0.1, 1)$ and $infinity$ (Table. \ref{tab:perc_loss}). Particularly, $\eta = 0$ means no perceptual loss, while $\eta = infinity$ means only perceptual loss is used. Similarly, $\gamma = (0, 0.001, 0.01, 0.1, 1)$ and $infinity$ were tested for the adversarial loss.

In our experiments, none of these values could lead to the best PSNR, SSIM and FID simultaneously, but when $\gamma = 0.001$ and $\eta = 0.006$, it performed well on both fidelity and perceptual evaluations.

Regarding the variations of adversarial loss, we also compared four popular GAN variations, which had been successfully used in SOTA SISR studies: vanilla GAN\cite{GAN, SRGAN}, RaGAN\cite{RaGAN, ESRGAN}, WGAN\cite{WGAN, MSGAN}, and WGANGP\cite{WGANGP, MSGAN}. They were trained with the same hyper-parameters from the same start point of a pre-trained generator, but with different designs of $\mathcal{L}_{adv}$. Vanilla GAN is using equation (\ref{eqt:GAN}); WGAN is using equation (\ref{eqt:WGAN}); WGANGP is using equation (\ref{eqt:WGANGP}); RaGAN is using:
\begin{equation}
\begin{split}
    & \mathcal{L}_{RaGAN} = \mathbb{E}_{I_{hr}}\left [ \log D(I_{hr}) - \mathbb{E}_{I_{lr}} \left [ \log D(G(I_{lr})) \right ] \right ] \\
    & + \mathbb{E}_{I_{lr}}\left [\log (D(G(I_{lr})))  - \mathbb{E}_{I_{hr}} \left [ \log D(I_{hr}) \right ]  \right ].
\end{split}
\end{equation}\label{eqt:RaGAN}

In our experiments (Fig. \ref{fig:gans}), WGANGP helped MIASSR achieve the best performance with all metrics. Therefore we choose WGANGP in MIASSR.

\subsection{Training Tricks}
We also found tricks of data processing that can affect model training with significant impacts on the final performance. First, to transfer the well-trained model on the OASIS dataset to new datasets could accelerate the training process, although it could not improve the final performance. Second, the networks, including both the generator and discriminator, must be initialised by a variety of uniform function (e.g., Kaiming-Uniform\cite{HeInit}) when they are trained from scratch. In our experiments, initialising networks with a normal distribution (e.g., Kaiming-Normal\cite{HeInit}) crashed the training process. Third, batch normalisation\cite{BN} should not be used, although it has succeeded in a wide range of image processing tasks. SRGAN, as the only method with batch normalisation, performed poorly in the patient-wise experiment (Table. \ref{tab:sota}). This is because the normalisation operation distorted the patient-wise contrast information during training, and led to poor performance on the test set. Finally, hyper-parameters such as patch size and image size can influence the final performance. In Fig. \ref{fig:sota}, although all methods tended to perform better with smaller magnification scales than larger ones in general, the best PSNR and SSIM scores always appeared with the x1.5 SR task. This is probably because 1.5 is the smallest SR scale to fit both the training patch size $[96 \times 96]$ and testing image shape $[144 \times 180]$.  

\begin{figurehere}
	\centering
	\includegraphics[width=3.2in]{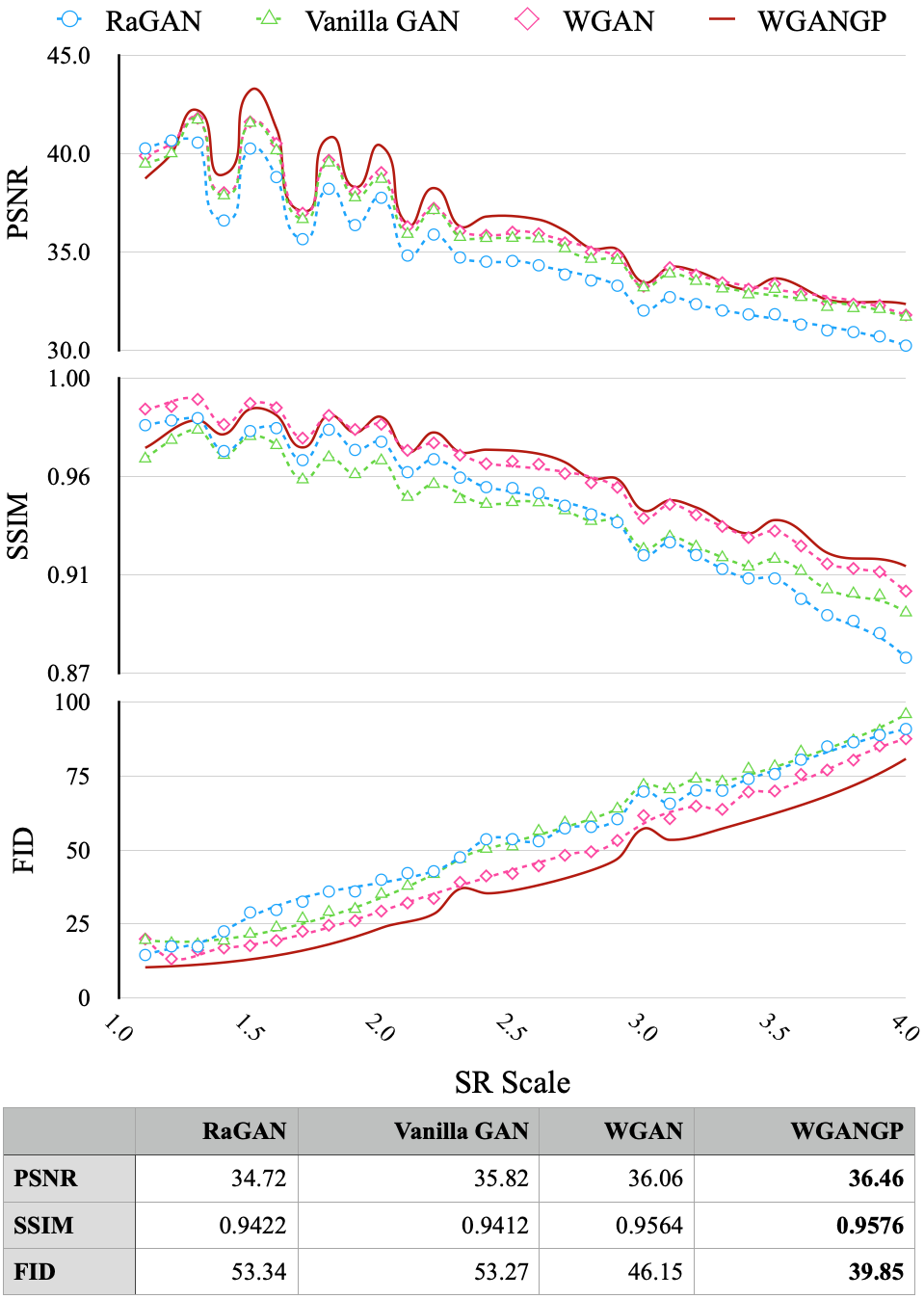}
	\caption{Comparing GAN variations in MIASSR. Higher PSNR and SSIM indicate better fidelity quality, while lower FID denotes more perceptual realistic images have been generated. WGANGP has not only achieved the best mean performance (the table at bottom) but also performed the best with almost all SR scales.}
	\label{fig:gans}
\end{figurehere}

\subsection{Future Work}
MIASSR can be extended in several potential directions. First, we believe that MIASSR can be extended to a wider range of medical image modalities and data types. In this work we focus on 2D slice SR because it is more general. In the future, to modify MIASSR with techniques such as 3D convolution and recurrent networks may extend it to 3D images \cite{3DPCA}, temporal scans \cite{LSTM}, and other 3D data such as mesh \cite{3DMesh} and point cloud \cite{3DCloud}. Secondly, we have had a taste of cross modality with naive transfer learning (ACDC and COVID-CT) and multi-task training (BraTS), and it would be interesting to investigate MIASSR further with more specific cross-modality applications. Thirdly, the ablation study of SR image generators and the adversarial learning helps us to understand each component of MIASSR well, and the conclusions and findings may also benefit other medical image analysis research studies, such as synthesis, reconstruction and segmentation. Finally, Fig. \ref{fig:num_paras} has reflected the balance between perception and distortion in either the sub-pixel based SR methods and meta-upscale module SR methods. Further research of this trade-off will be useful in both research and clinical application development. Furthermore, to find a more straightforward measurement of medical images in SR tasks is also desired. For example, instead of measuring the quality of generated images, to evaluate the images in specific tasks such as AD diagnosis \cite{ADDiagnosis} and Parkinson disease identification \cite{Parkinson1} could be a potential way.

\section{Conclusions}
In this paper, we have proposed MIASSR for medical image super-resolution tasks with arbitrary scales. To our best knowledge, this is the first attempt to develop a meta-learning scheme for this problem. We have first reduced the model size (only 26\% parameters as compared to existing meta-SR methods) by using a lite EDSR model as the LR image feature extractor and achieved comparable fidelity quality of SR images with SOTA methods. Moreover, we have introduced GANs to meta-learning to improve the perceptual quality of the generated images. The proposed method has obtained good practical applicability. In our experiments, we have successfully applied MIASSR with T1-weighted brain MR images and multi-modal brain MR scans. Furthermore, with transfer learning, the pre-trained model on brain images has been fine-tuned to cardiac MR scans and CT scans. We have also discussed our findings and understandings of model architecture design, training tricks, and adversarial learning in the comparison studies.

\section{Acknowledgements}
This work was supported in part by China Scholarship Council (grant No.201708060173), in part by the British Heart Foundation [PG/16/78/32402], in part by the European Research Council Innovative Medicines Initiative on Development of Therapeutics and Diagnostics Combatting Coronavirus Infections Award 'DRAGON: rapiD and secuRe AI imaging based diaGnosis, stratification, fOllow-up, and preparedness for coronavirus paNdemics' [H2020-JTI-IMI2 101005122], in part by the AI for Health Imaging Award 'CHAIMELEON: Accelerating the Lab to Market Transition of AI Tools for Cancer Management' [H2020-SC1-FA-DTS-2019-1 952172], in part by the Hangzhou Economic and Technological Development Area Strategical Grant [Imperial Institute of Advanced Technology], in part by the GO-DS21 EU grant proposal. The authors would like to thank Ben Day for constructive criticism of the manuscript.

\bibliographystyle{ws-ijns}
\bibliography{MIASSR_IJNS.bib}

\end{multicols}

\appendix{Extend MIASSR to various medical image modalities}

\begin{figure*}[h]
	\centering
	\includegraphics[width=\textwidth]{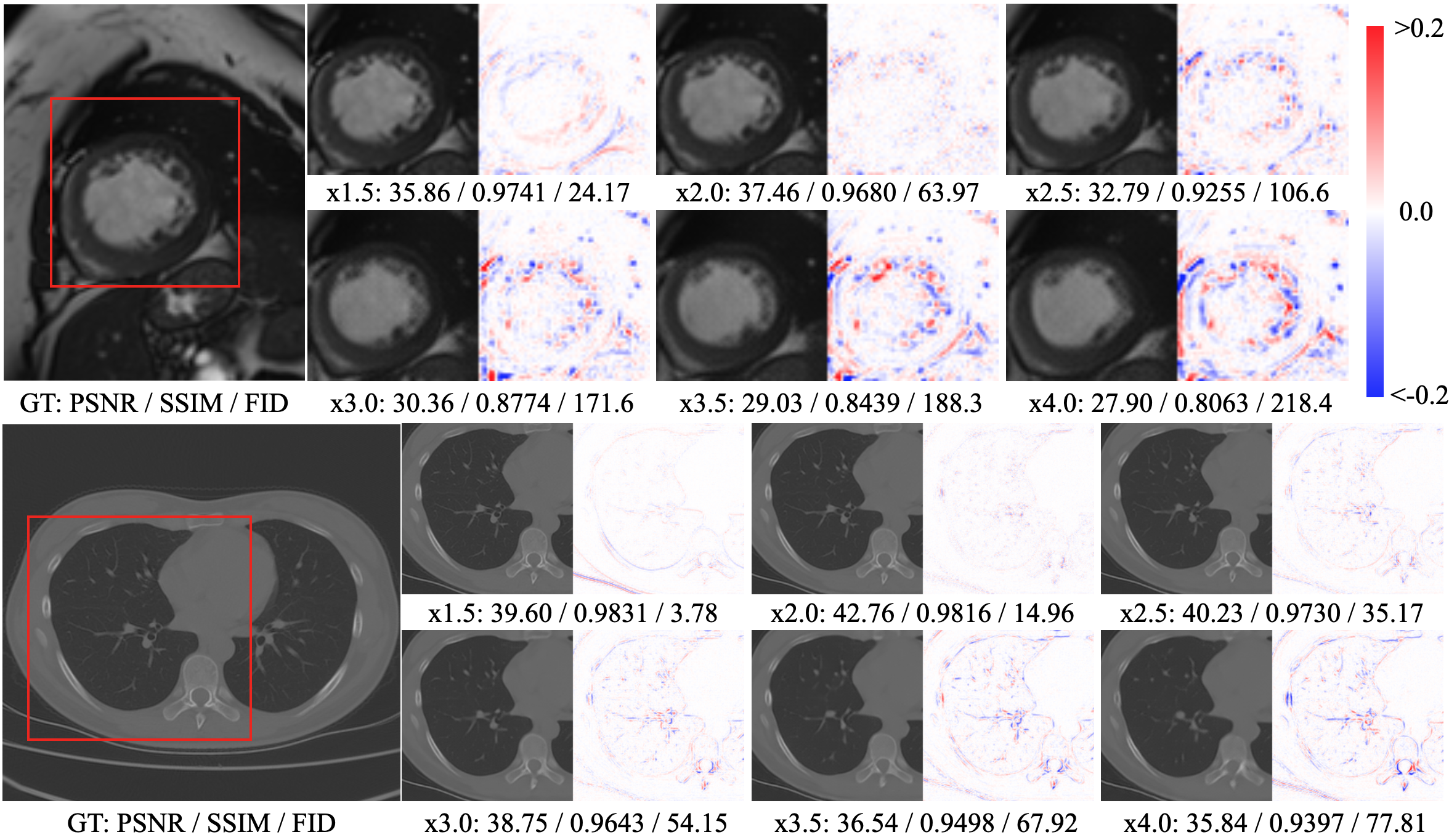}
	\caption{With transfer learning, MIASSR can be applied to new medical image modalities conveniently and effectively. We have extended it to cardiac MR images (top), and CT images of COVID-CT patients (bottom). Ground truth image is plotted on the left, while SR images with various magnification scales and the pixel-wise errors are on the right.}
	\label{fig:acdc_covid_example}
\end{figure*}

\begin{figure*}[p]
	\centering
	\includegraphics[width=\textwidth]{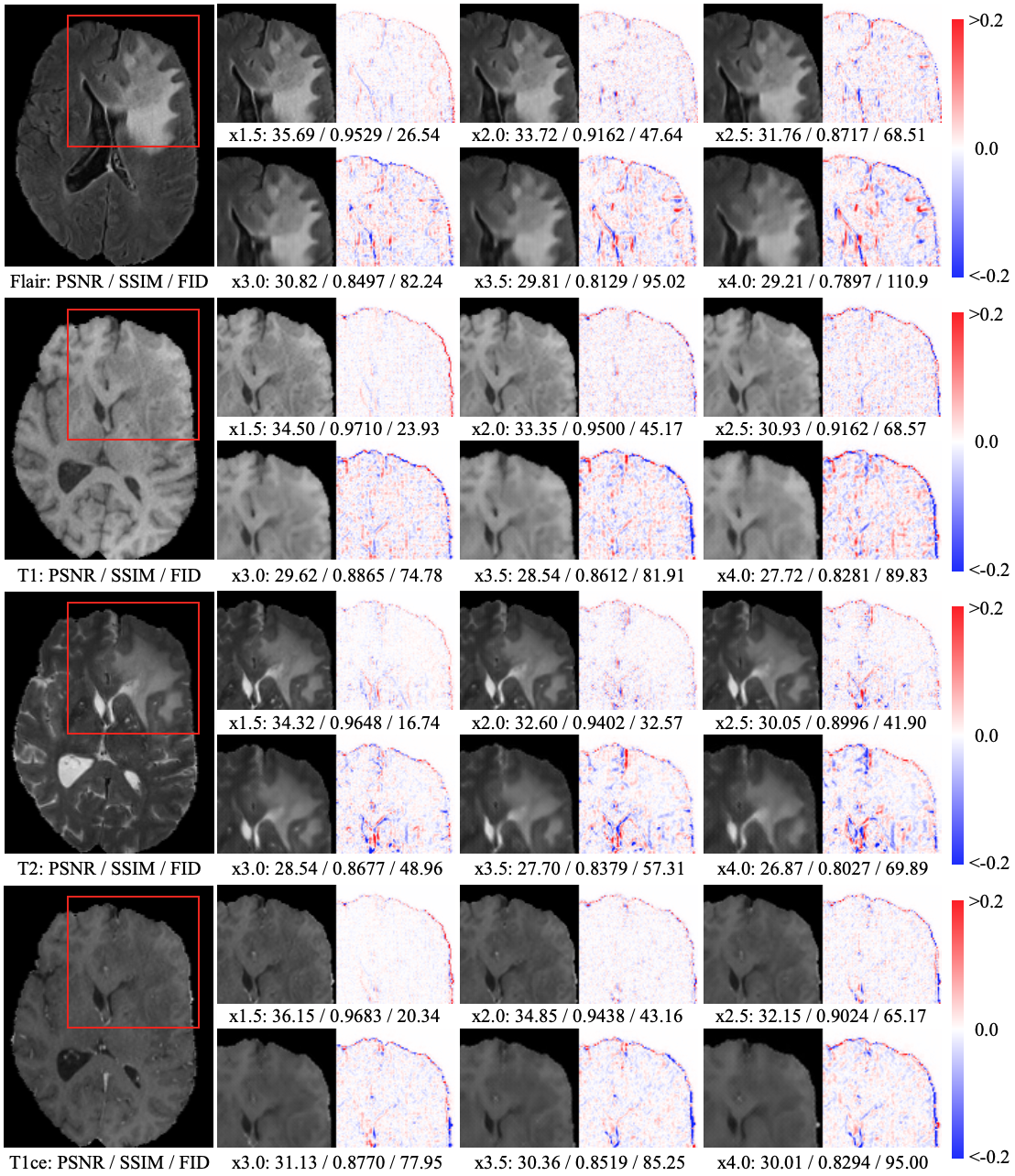}
	\caption{MIASSR has successfully worked with multi-modal brain MR image dataset BraTS. All images are converted to [0, 1]. Differences between SR images and ground truth images are rendered.}
	\label{fig:brats_example}
\end{figure*}

\end{document}